\documentstyle[12pt]{article}
\input epsf

\begin{document}

\begin{center}
\Large{{\bf A study of the angular size - redshift relation for 
models in which $\Lambda$ decays as the energy density}} 
\end{center}

\vspace{1cm}
\begin{center}
 R. G. Vishwakarma

\vspace{.6cm}
{\sf IUCAA, Post Bag 4, Ganeshkhind, Pune 411 007, India\\
E-mail: vishwa@iucaa.ernet.in}
\end{center}

\vspace{2cm}
\noindent
{\bf Abstract.}
By modifying the Chen and Wu ansatz, we have investigated some 
Friedmann models in which $\Lambda$ varies as $\rho$. In order 
to test the consistency of the models with observations, we study 
the angular size - redshift relation for 256 ultracompact radio 
sources selected by Jackson and Dodgson. The angular sizes of 
these sources were determined by using very long-baseline 
interferometry in order to avoid any evolutionary effects. 
The models fit the data very well and require an accelerating 
universe with a positive cosmological constant. Open, flat and 
closed models are almost equally probable, though the open model 
provides a comparatively better fit to the data. The models are 
found to have intermediate density and  imply the existence of 
dark matter, though not as much as in the canonical Einstein-de 
Sitter model.

\vspace{.5cm}

\vspace{1cm}

\noindent
{\bf 1. Introduction}

\noindent
The cosmological constant $\Lambda$, which was originally invoked by
Einstein to obtain a static solution of his field equations but was
subsequently rejected by him after the realization that the
universe is expanding, has fallen in and out of fashion several times.
However recent observations by Perlmutter et al (1999)
and Riess et al (1998) strongly favour a significant and positive 
$\Lambda$. Their findings arise from the study of more than 50 
type Ia supernovae with redshifts in the range $0.16\leq z\leq $ 0.83 
and suggest Friedmann models with negative pressure-matter such as a 
cosmological constant, domain walls or cosmic strings (Vilenkin 1985,
Garnavich et al 1998). The main conclusion of these
works is that the expansion of the universe is accelerating.

Although a cosmic acceleration can also be accounted for by 
invoking inhomogeneity (though at the cost of the Cosmological 
Principle) (Pascual-Sanchez 1999, Dabrowski 1999),
postulating a $\Lambda$-dominated model solves a lot of problems
at once. The cosmological constant supplies the ``missing matter" 
required to make $\Omega_{\mbox{{\scriptsize tot}}}=1$ (as 
suggested by the inflationary models, though on the basis of 
little observational evidence).  It modifies CDM
by putting more power on large scales, as is compatible with the 
CMB anisotropy limits. It also removes the inconsistency between 
the age of the universe and that of the globular clusters for 
larger values of the Hubble parameter $H_0$ (the subscript zero 
denotes the value of the quantity at the present epoch). 

However, even with this significant value of $\Lambda$,  we 
obviously face the problem that the upper limit of $\Lambda$ from 
observations is 120 orders of magnitude below the 
value for the vacuum energy density predicted by quantum field 
theory (Weinberg 1989; Carroll, Press and Turner 1992). (It is
customary to associate a positive cosmological constant $\Lambda$ 
with a vacuum density $\rho_v \equiv \Lambda/8\pi G$.)

Different types of solutions have been proposed to solve this 
problem and these can be classified into two categories. The first 
one, advocated by particle physicists, implements 
some kind of adjustment mechanism, for instance, a counter term in 
the Lagrangian which goes away with the effective cosmological
constant. However, as remarked by Zee (1985), apart from the 
fine tuning of the parameters required by this method, there is no 
known symmetry which would guarantee the effective $\Lambda$-term 
being zero. The second approach, advocated by  general relativists, 
considers $\Lambda$ as a dynamic variable. The link between the 
two categories is generally considered by studying different classes 
of scalar fields (Ratra and Peeble 1988).

The second approach, which is essentially phenomenological in
nature, has been extensively investigated in the past few years. 
It argues that, due to the coupling of the dynamic degree of 
freedom with the matter fields of the universe, $\Lambda$ relaxes 
to its present small value through the expansion of the universe 
and the creation of photons. From this point of view, the cosmological
constant is small because the universe is old.

Several ansatzes have been proposed in which the $\Lambda$ term 
decays with time (Gasperini 1987, Freese et al 1987, Ozer and 
Taha 1987, Gasperini 1988, Peebles and Ratra 1988, Chen and Wu
1990, Abdussattar and Vishwakarma 1996, Gariel and Denmat 1999). 
Of special interest is the ansatz $\Lambda \propto S^{-2}$
($S$ being the scale factor of the Robertson Walker metric) by 
Chen and Wu, which has been considered/modified by several authors
(Abdel-Rahman 1992; Carvalho, Lima and Waga 1992; Waga 1993; 
Silveira and Waga 1994). Through dimensional considerations, in 
the spirit of quantum cosmology, Chen and Wu argued that  one can 
always write the vacuum energy density as $M_{\mbox{{\scriptsize Pl}
}}^4$ times a dimensionless quantity where $M_{\mbox{{\scriptsize Pl}
}}=(\hbar c/G)^{1/2}$ is the Planck mass. They therefore considered

\begin{equation}
\Lambda \propto \frac{1}{l_{\mbox{{\scriptsize Pl}}}^2}\left[\frac{
l_{\mbox{{\scriptsize Pl}}}}{S}\right]^n,
\end{equation}
where $l_{\mbox{{\scriptsize Pl}}}=(G\hbar/c^3)^{1/2}$ is the Planck 
length. They noted that if one estimates $S_0$ by $c t_0$, then 
$n\leq1$ or $n \geq3$ would lead to either too big or too 
small values of $\Lambda_0$ compared to the observed limit. 
Consequently one needs to put $n=2$, which gives the right 
value.

Carvalho et al (1992) realized that (1) is not the only possible 
dynamic law for $\Lambda$ given by the Chen and Wu ansatz. One may, 
for example, also write

\begin{equation}
\Lambda \propto \frac{1}{l_{\mbox{{\scriptsize Pl}}}^2}\left[\frac{
t_{\mbox{{\scriptsize Pl}}}}{t_H}\right]^n,
\end{equation}
where $t_{\mbox{{\scriptsize Pl}}}$ and  $t_{\mbox{{\scriptsize H}}}
\equiv H^{-1}$ are Planck and Hubble times respectively. With the 
same argument as stated above, one finds $\Lambda \propto H^2$.

Thus the Chen and Wu ansatz was generalized by Carvalho et al by
considering $\Lambda=\alpha S^{-2}+\beta H^2$  and later on by Waga 
(1993) by considering $\Lambda=\alpha S^{-2}+\beta H^2+\gamma$,  
where $\alpha, ~ \beta $ and $\gamma$ are  adjustable dimensionless 
parameters.

In this paper, we modify the Chen and Wu ansatz further.
Since a non-zero $\Lambda$ might be thought of as producing 
significant non-gravitational long range forces in the evolution of the 
universe and consequently influencing the large-scale dynamics, 
it is natural to hope that the sought dynamical law for $\Lambda$ 
might also depend upon some cosmological parameter related
to the material content in the universe. Moreover, there is still
room for any dimensionless parameter in the Chen and
Wu ansatz. One parameter of interest is the density parameter 
 $\Omega$, which is the density of the matter in units of
the critical density, i.e., $\Omega\equiv \rho/\rho_c =8\pi G 
\rho/3H^2$. We thus consider the ansatz 

\begin{equation}
\Lambda=n~\Omega~H^2,
\end{equation}  
where $n$ is a new dimensionless cosmological parameter to be 
determined from the observations. We here do
not consider the $S^{-2}$-dependence because, in view of the present
estimate of $\Lambda$ being of order of $H_0^2$, an $H^2$-dependence
seems more natural if $\Lambda$ is to relax to its present estimate 
due to the expansion of the universe. 

Equation (3), which can alternatively be written as $\Lambda =(8n\pi
G/3)\rho$ or $\Omega_\Lambda=(n/3)\Omega$, puts $\Lambda$ on the
same footing as the energy density $\rho$ and this corroborates the 
Machian view. (Here $\Omega_\Lambda \equiv \rho_v/\rho_c =\Lambda/3H^2$ 
denotes the energy density of vacuum in units of the critical density.)

Incidentally we note that if one estimates the present ``radius" of 
the universe by $S_0 \approx ct_0 \approx cH_0^{-1}$, then the 
present gravitational attraction $4\pi cG\rho_0/3H_0$ roughly 
balances the cosmic repulsion $\Lambda_0 c H_0^{-1}$ provided 
$\Lambda_0\approx \Omega_0 H_0^2$. 

In this paper we investigate cosmological models based on the
dynamical law (3), in the framework of general relativity, and study
their observational implications. The paper is organized as follows: 
The basic equations describing the models are presented in Section 2.
In Section 3, we study the angular size - redshift relation in the
models, this providing one of the classical tests of the 
observational cosmology. For this purpose, we consider the data set 
of 256 ultra compact radio sources (with redshifts in 16 bins 
in the range 0.5 - 3.8) from Jackson and Dodgson (1997). Our results 
are discussed and compared with previous works in Section 4.

\vspace{.5cm}
\noindent
{\bf 2. The models and the field equations}

\noindent
In order to introduce the ansatz (3) and compare our conclusions 
with earlier works, we here consider the 
Friedmann-Lemaitre-Robertson-Walker models. For a perfect fluid, 
the models are fully specified by the scale factor $S(t)$ and the 
curvature index $k \in \{-1,0,1\}$ of the spatial hypersurfaces 
$t=$ constant. Denoting the derivative with respect to the cosmic 
time $t$ by a dot and using units with $c=1$, the models are 
governed by

\noindent
the Raychaudhuri  equation:

\begin{equation}
-\frac{\ddot S}{S}=\frac{4\pi G}{3}\left(\rho+3p\right)-\frac
{\Lambda}{3}
\end{equation}

\noindent
and the Friedmann equation:

\begin{equation}
\frac{\dot S^2}{S^2}+\frac{k}{S^2}=\frac{8\pi G}{3}\rho+\frac
{\Lambda}{3}.
\end{equation}
We also assume that $p=w\rho$ is the equation of state of the 
perfect fluid matter. Causality then requires $-1\leq w\leq1$. 
With the help of (3), equation (4) reduces to 

\begin{equation}
-\frac{\ddot S}{S}=\frac{4\pi G}{3}\left(1+3w-\frac{2n}{3}
\right)\rho,
\end{equation}
implying that inflationary solutions (i.e., solutions with 
negative
deceleration parameter $q\equiv -S \ddot S/\dot S^2 $) are 
possible for $n > 3(1+3w)/2$. We also note that the models
are open ($k<0$), flat ($k=0$) or closed ($k>0$) according as 
$\Omega \frac{<}{>} 3/(n+3)$, respectively, provided
$n>-3$. For $n\leq -3$, the non-static model is always open.

The Einstein field equations, via the Bianchi identities, imply

\begin{equation}
\left(R^{ij}-\frac{1}{2} R g^{ij}\right)_{;j}=0=-8\pi G
\left(T^{ij}-\frac{\Lambda}{8\pi G}g^{ij}\right)_{;j}
\end{equation}
leading to the energy conservation equation

\begin{equation}
\dot\rho+3(1+w)\rho\frac{\dot S}{S}+\frac{\dot\Lambda}{8\pi G}=0.
\end{equation}
By using equation (3), this leads to

\begin{equation}
\rho=CS^{-9(1+w)/(n+3)}, ~ ~ n\neq -3, ~ ~  C=\mbox{constant} >0.
\end{equation}
For $n=-3$, the dynamic model yields $(1+w)\rho=0$ with $S=t$ and $k=-1$.
This leads to the following two cases. (i) Milne's model:
$\rho=p=\Lambda=0$, (ii) $w=-1$ with the only constraint on 
$\rho$ and $\Lambda$ coming from equation (3) leaving their
functional form otherwise free. However, $w=-1$ just corresponds 
to a second cosmological constant and will not be considered 
further. With the help of (9), the Friedmann  equation (5) can 
be integrated to give

\begin{equation}
t=\int_{S_i}^{S}\left[(n+3) \frac{8\pi GC}{9}
y^{(2n-3-9w)/(n+3)}-k\right]^{-1/2}dy, ~ ~ n\neq -3,
\end{equation}
where we have chosen the origin of time at $S=S_i ~ (\neq 0)$
with $\dot S=0$ (occurring at the first time) in the models 
without big bang and at $S_i=0$ in those with big bang. 
Equation (10) implies that $n>-3$ for $k=0$ or 1. In the case 
$k=1$, $S$ is restricted by 

\begin{equation}
S^{(2n-3-9w)/(n+3)}\geq 9/8\pi (n+3)GC.
\end{equation}

The integral in equation (10) can easily be
evaluated for a general value of $n$ if $k=0$. For $k
\ne 0$, it can still be evaluated for particular values of 
$n$ and we describe one such solution below. However numerical
integration is possible for any particular value of $n$.

For $ k=0$, equation (10) yields

\begin{equation}
S=\left[3(1+w)\sqrt{\frac{2\pi GC}{(n+3)}} ~
t\right]^{(2/9)(n+3)/(1+w)}.
\end{equation}
Thus the scale factor $S$ evolves as $S\propto t^{(n+3)/6}$
in the radiation-dominated (RD) era and $S\propto t^{2(n+3)/9}$ 
in the matter-dominated (MD) era, which is a simple 
generalisation of the canonical Einstein-de Sitter solution. 
The modified expression for the energy density,

\begin{equation}
\rho=\frac{n+3}{18(1+w)^2\pi G} t^{-2} 
\end{equation}
in the present case, has the same time-variation as 

\[
\rho= \frac{1}{6\pi G(1+w)^2} t^{-2} 
\]
in the Einsten-de Sitter model.
The expression for $\Lambda$ in the present case becomes 

\begin{equation}
\Lambda=\frac{4n(n+3)}{27(1+w)^2} t^{-2},
\end{equation}
which matches with the natural dimensions of $\Lambda$.

\vspace{1cm}

\noindent
{\bf Solutions for $k\ne 0:$}
 The solutions with $n=1$ are as follows.

\vspace{.5cm}
\noindent
\underline{\large {$k=1, ~ w=1/3$}:}

\begin{equation}
t=\sqrt{aS-S^2} +\frac{a}{2}\sin^{-1}(2aS-1)-\frac{3\pi a}{4},
\end{equation}
where $a=(32\pi GC)/9$.

\vspace{.5cm}

\noindent
{\bf \underline{\large{$k=1, ~ w=0$:}}}

\[
t=a^4\{-\cos x \sin^7 x-\frac{7}{6}\cos x \sin^5 x-
\frac{35}{24}\cos x
\sin^3 x
\]
\begin{equation}
-\frac{35}{16}\cos x\sin x+\frac{35}{16}x\}, 
\end{equation}
with 

\begin{equation}
a\sin^2 x=S^{1/4}.
\end{equation}

\vspace{.5cm}
\noindent
\underline{\large{{\bf $k=-1, ~ w=1/3$}}}:

\begin{equation}
t=\sqrt{aS+S^2}-\frac{a}{2}\cosh^{-1} (2aS+1).
\end{equation}

\noindent
\underline{\large{{\bf $k=-1, ~ w=0$}}}:

\[
t=a^4\{\cos^{-8}x
\sin^7x-\frac{7}{6}\cos^{-6}x\sin^5x+\frac{35}{24}
\cos^{-4}x \sin^3x-\frac{35}{16}\cos^{-2}x\sin x
\]
\begin{equation}
+\frac{35}{16}\ln (\sec x+\tan x)\},
\end{equation}
with 

\begin{equation}
a\tan^2 x=S^{1/4}.
\end{equation}

One can also approximate the solutions for a general $n$ if 
there is no significant variation in the value of $\Omega$ 
compared to that in $H$ during the expansion of the universe. 
(This assumption is trivially satisfied in models with 
$n=3(1+3w)/2$ and obviously in flat models also.)
By using equation (3), the Raychaudhuri equation (4) can be 
written as 

\begin{equation}
\dot H+\left(\{3+9w-2n\}\frac{\Omega}{6}+1\right)H^2=0.
\end{equation}
If the variation in $\Omega$ during a time interval is not 
significant compared to that in $H$, one can approximate the 
solution of equation (21) as

\begin{equation}
t\approx mH^{-1},
\end{equation}
where $m=6/[6-(2n-3-9w)\Omega]$.
This gives an approximate age of the universe as $t_0\approx
mH_0^{-1}$. Hence $t_0>H_0^{-1}$ for $n>1.5$. The corresponding 
approximate evolution of $S$ turns out to be 

\begin{equation}
S\propto t^m.
\end{equation}
Although we shall not use these approximate solutions given by 
equations (22) and (23) in further analysis, we shall see in 
section 4 that the estimates of $t_0$ calculated from these 
equations are very close to those obtained from equation (10).

In order to compare the models with observations and determine the
relative contributions of the different cosmological parameters
at the present epoch, it is necessary to recast equations (4) and 
(5) in the following forms:

\begin{equation}
2[q_0 + \Omega_{\Lambda 0}]=\Omega_{0},
\end{equation}

\begin{equation}
1+\Omega_{k0} = \Omega_{0} + \Omega_{\Lambda 0}.
\end{equation}
Here $\Omega_k \equiv k/S^2 H^2$ is defined as the curvature
parameter. Now the age of the universe follows from equation (10) 
as

\begin{equation}
H_0 t_0=\int_{\phi_i}^1
\left[\left(1+\frac{n}{3}\right)\Omega_{0} \phi^{(2n-3)/(n+3)} -
\Omega_{k0}\right]^{-1/2} d\phi,
\end{equation}
where $\phi=S/S_0$ and $\phi_i=S_i/S_0$.

\newpage
\noindent
{\bf 3. Comparison of the models with observations }

\noindent
We now want to study the consistency of our models with 
observations. We know that as we look back from our position 
at $r=0$ and $t=t_0$ to some object at a radial coordinate 
distance $r_1$, we are also looking back to some time 
$t_1<t_0$ and some expansion factor $S_1=S(t_1)<S_0$. Note,
however, that neither $r_1,  t_1$, nor $S_1$ are directly 
measurable; what are measured are physical properties
like redshift, proper or apparent angular sizes, velocities, 
luminosities etc. Since the various cosmic distance measures 
depend sensitively on the parameters of the models, the 
physical properties of the distant objects are also influenced 
by these parameters. In particular, the dependence of the
angular size $\Theta$ of a standard measuring rod upon redshift
$z$ depends upon these parameters. For this reason,
the $\Theta$ - $z$ relation was proposed as a potential test for
cosmological models by Hoyle (1959). The original expectation 
that the test would be able to distinguish between the 
geometries of the various cosmological models, was not fulfilled 
because of the intrinsic scatter and evolution of the sources. 
However, Kellermann (1993) argued that the evolutionary effects 
could be controlled by choosing a sample of ultra compact radio 
sources, with angular sizes of the order of a few milliarcseconds 
measured by very long-base line interferometry (VLBI). These 
sources, being short-lived and  deeply embedded inside the 
galactic nuclei, are expected to be free from evolution on a 
cosmological time scale and thus comprise a set of standard
objects, at least in a statistical sense. He used a sample 
of 79 such sources and showed that a credible $\Theta$ - $z$ 
relation emerged. His results were, however, limited to showing that the 
corresponding $\Theta$ - $z$ diagram favoured an Einstein-de 
Sitter model (with $\Omega_0 =1$), thus implying the existence 
of a large amount of cold dark matter. Jackson and Dodgson (1996) 
extended this work by including a $\Lambda$ term and showed that 
Kellermann's data could also fit  a low density, highly 
decelerating model with large negative $\Lambda$. 

Later on, more extensive exercise was carried out on
a bigger sample of 256 ultra compact sources by Jackson and 
Dodgson (1997). They then concluded that the canonical 
inflationary cold dark matter model ($\Omega_0=1, ~ \Omega_{
\Lambda }=0$) was ruled out at the 98.5 percent confidence 
level and that a low-density Friedmann model with either sign of 
$\Lambda$ was favoured. 

The original data set for the ultra compact sources used by 
Jackson and Dodgson was compiled by Gurvits (1994). The full 
sample included 337 sources, out of which Jackson and 
Dodgson selected the sources with $z$ in the range 0.5 to 3.8 
for reasons discussed in their paper. These sources, 256 in 
number, were binned into 16 redshift bins, each bin containing 
16 sources. Recently the compilation of Jackson and Dodgson has 
been used by Banerjee and Narlikar (1999) in the quasi-steady
state cosmology. We also use this data set to study the 
$\Theta$ - $z$ relation in the present models and check its 
consistency with observations. For this purpose,
we now derive the $\Theta$ - $z$ relation in the models.

If the observer at $r=0$ and $t=t_0$ receives the light from a 
source at a radial distance $r_1$ with redshift $z$, the Hoyle's 
formula gives the apparent angular size $\Theta$ of the source as

\begin{equation}
\Theta=\frac{d(1+z)}{r_1 S_0},
\end{equation}
where $d$ is the known (or assumed) proper size of the source 
and the coordinate radius $r_1$ is given by

\begin{equation}
\eta (r_1) =\int_{S_0/(1+z)}^{S_0} \frac{dS}{S\dot S}
\end{equation}
with

\[
\eta (r_1)=\sin^{-1} r_1, ~ ~ ~ ~ k=1
\]
\[~ ~ ~ ~  ~ ~ ~ =r_1, ~ ~ ~  ~ ~  ~ ~ ~ ~ ~ ~ ~ k=0
\]
\begin{equation}
~ ~ ~ ~ ~ ~ ~ ~ ~ ~ ~ =\sinh^{-1} r_1, ~ ~ ~ k=-1.
\end{equation}
Using (3), (5) and (9), equation (28) reduces to

\begin{equation}
\eta (r_1) = \frac{1}{S_0 H_0} \int_1^{1+z}
\left[\left(1+\frac{n}{3}\right)\Omega_{0} \psi^{9/(n+3)} -
\Omega_{k0} \psi^2 \right]^{-1/2} d\psi,
\end{equation}
where $\psi=S_0/S$. We also note that 

\begin{equation}
S_0=\frac{1}{H_0}\sqrt{\frac{k}{\Omega_{k0}}}, ~ ~ ~ ~ k\neq 0
\end{equation}
and equations (3) and (25) give 

\begin{equation}
n = \frac{3}{\Omega_{0}}\left(1+\Omega_{k0}\right)-3.
\end{equation}

It is clear from equations (27) and (29) - (32) that, once we fix
$\Omega_{k0}$ and
$\Omega_0$, the theoretical $\Theta$ - $z$ relation can be 
completely worked out for given $d$ and $H_0$. We assume 
$H_0 =$ 65 km/s/Mpc and $d = 10$ pc (this does not have any 
consequences as there is still quite uncertainty in the value of
$H_0$) and calculate the theoretical $\Theta(z)$ at the mean bin 
redshifts for a range of parameters $\Omega_0$ and $\Omega_{k0}$. 
Using the observed values of $\Theta_i$ and the same standard 
errors $\sigma_i$ of the $i$th redshift bin as used by Jackson 
and Dodgson, we compute $\chi^2$ according to

\begin{equation}
\chi^2=\sum_{i=1}^{16} \left[\frac{\Theta_i -
\Theta(z_i)}{\sigma_i}\right]^2.
\end{equation}
This has 16 degrees of freedom as there are no constraints on 
the parameter space. We have performed an extensive investigation 
in order to find best-fit parameters. For this purpose, we  have 
tried to minimise $\chi^2$ with respect to $\Omega_{k0}$ and 
$\Omega_0$ and find the best-fitting solution as 
$\Omega_{k0}=-0.22$ and $\Omega_0=0.49$, which is an open model. 
However, flat and closed models also fit the data quite 
satisfactorily. This has been shown through some examples and the
results have been plotted in Figures 1 and 2.

\newpage

\centerline{{\epsfxsize=14cm {\epsfbox[50 250 550 550]{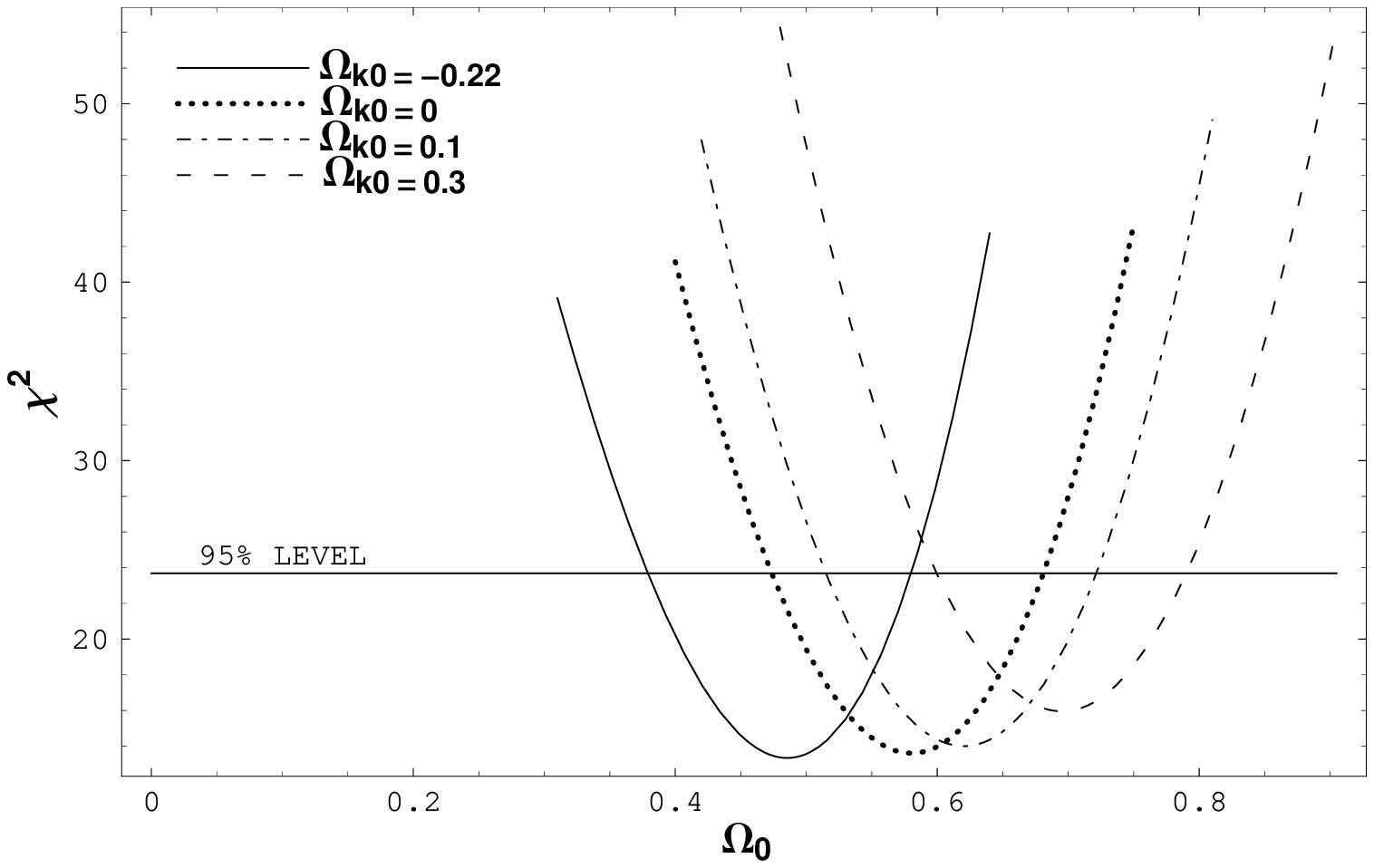}}}}

\vspace{.7cm}

{\bf Figure 1.} $\chi^2$ plotted against $\Omega_0$ for 4
typical models. The minimum

value of $\chi^2$ decreases for lower density models.

\vspace{1cm}

\centerline{{\epsfxsize=14cm {\epsfbox[50 250 550 550]{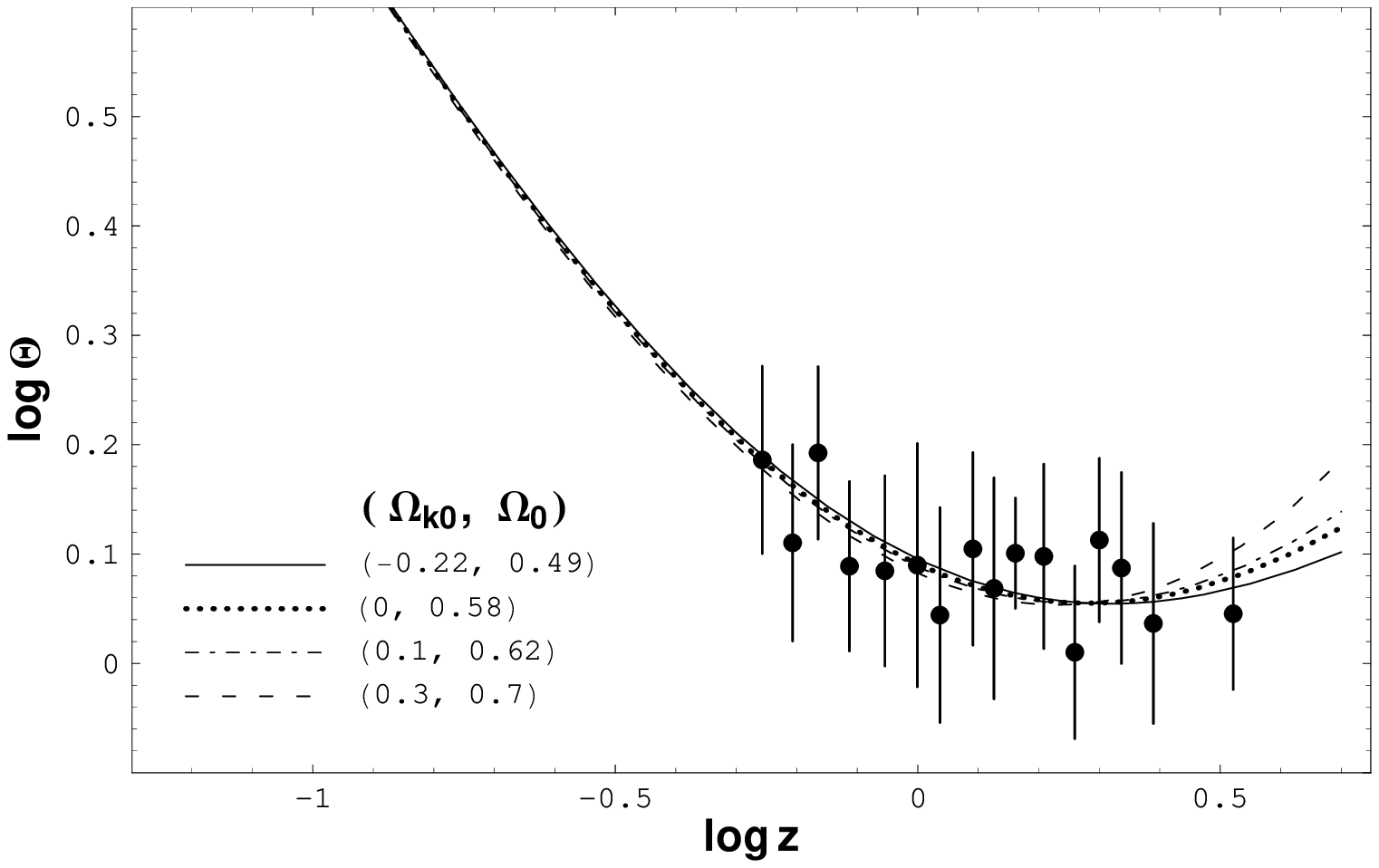}}}}

\vspace{.5cm}
{\bf Figure 2.} Best-fitting $\log \Theta$/$\log z$ curves
for 4 models, to be compared

with Jackson and Dodgson' data points.

\vspace{1cm}
\noindent
{\bf 4. Interpretation and conclusion}

\noindent
Figure 1 illustrates four typical cases of $\Omega_{k0}$: for each
$\Omega_{k0}$, we have plotted the values of $\chi^2$ against 
$\Omega_0$. The 95 percent line (with 14 degrees of freedom) 
intersects each curve and leaves sufficient portions of the curves 
below it, 
which shows a reasonably good fit. The minimum value of $\chi^2$ 
decreases for lower density universes. Moreover, in accordance 
with the standard model, we have lower density for the open model. 
However, we also note that, contrary to the claims of Jackson and 
Dodgson (1997), the closing of the universe with 
$\Omega_{\Lambda 0}$ does not abolish the need to postulate 
nonbaryonic matter, though the density is not as high as in
Kellermann's model. Although the test is not conclusive in 
determining the curvature signature $k$, leaving all three signs 
almost equally probable, the fit of the data with the open model 
is excellent, with a probability exceeding 50 percent! The 
corresponding probability for the other optimum case 
($\Omega_{k0}=0.3$) is about 30 percent which also represents 
a good fit. The other two cases lie in  between. The predicted 
age limit in these optimum cases is about $(1 - 1.4)H_0^{-1}$. 
The formal best-fitting values are shown in Table 1.

\vspace{1cm}

{\bf Table 1:} Best-fitting values of the different parameters
for four typical

cases of $\Omega_{k0}$.

\vspace{.5cm}
\begin{center}
\begin{tabular}{|l|l|l|l|l|l|}
\hline\hline
\multicolumn{1}{|c|}{ $\Omega_{k0}$}&
\multicolumn{1}{c|}{$\Omega_0$}&
 ~ $\Omega_{\Lambda0}$&
~ ~ $ n$&
 ~ ~ $\chi^2$&
\multicolumn{1}{c|}{~$ H_0 t_0$}\\
\hline
~-0.22 ~& ~ 0.49 ~ & ~ 0.29 ~ & ~ 1.79
 ~ & ~ 13.34 ~ & ~ 1.05 ~\\

~ 0 ~ & ~ 0.58 ~ & ~ 0.42 ~ & ~ 2.17 
~ & ~ 13.61 ~ & ~ 1.15 ~\\

~ 0.1 ~ & ~ 0.62 ~ & ~ 0.48 ~ & ~ 2.32 
~ & ~ 14.00 ~ & ~ 1.22 ~\\

~ 0.3 ~ & ~ 0.70 ~ & ~ 0.60 ~ & ~ 2.61 
~ & ~ 15.96 ~ & ~ 1.41 ~\\
\hline
\end{tabular}
\end{center}

\vspace{.8cm}

It is, therefore, clear  from Table 1 that the data favour an
accelerating universe with a positive $\Lambda$
which is interesting in view of the results from the supernovae 
data. However, it may be noted that a decelerating 
universe also cannot be ruled out completely. We mention, for 
example, a model $\Omega_0=0.42$ and $\Omega_{\Lambda0}=-0.02$
with $d=8.9$ pc which has a fairly good fit ($\chi^2=15.39$) to 
the data. However, this model 
is not interesting on the grounds of age considerations. It may 
be noted that if we use the approximate solution given by 
equations (22) and (23) to calculate $H_0 t_0$, the estimates
never differ by more than 4.5 percent from those shown in Table 1.

In Figure 2, we have plotted the theoretical $\Theta$ - $z$ 
curves for the four models discussed above and compared them with 
the Jackson and Dodgson' data points. The figure shows that the 
models fit the data very well.

\vspace{.3cm}

\noindent
{\bf Acknowledgements}

\noindent
The author thanks Professor J V Narlikar for fruitful and 
inspiring discussions. The author also gratefully acknowledges 
IUCAA and MRI for hospitality where this work was done. 
Thanks are also due to the referees for useful comments which 
helped in improving the paper.


\vspace{.5cm}
\noindent
{\bf References:}

\noindent
Abdel-Rahaman A-M M 1992 Phys. Rev. D {\bf 45} 3492\\
Abdussattar and Vishwakarma R G 1996 Pramana - J. Phys {\bf 47} 41\\
Banerjee S K and Narlikar J V 1999 MNRAS {\bf 307} 73\\ 
Carroll S M Press W H and Turner E L 1992 Ann. Rev. Astron.

\hspace{.5cm} Astrophys. {\bf 30} 499

\noindent
Carvalho J C Lima J A S and Waga I 1992 Phys. Rev. D {\bf 46} 2404\\
Chen W and Wu Y S 1990 Phys. Rev. D {\bf 41} 695\\
Dabrowski M P 1999 Preprint gr-qc/9905083\\
Freese K Adams F C Friemann J A and Mottolla E 1987 Nucl.

\hspace{.5cm} Phys. B {\bf 287} 797

\noindent
Gariel J and Le Denmat G 1999 Class. Quant. Gravit. {\bf 16} 149\\
Garnavich et al 1998 Astrophys. J. {\bf 509} 74\\
Gasperini M 1987 Phys. Lett. B {\bf 194} 347\\
\rule{1.2cm}{0.1mm} 1988 Class. Quant. Gravit. {\bf 5} 521\\
Gurvits L I 1994 Astrophys. J. {\bf 425} 442\\
Hoyle F 1959 in Bracewell R N ed. IAU Symp. No. 9 Paris Sympo.

\hspace{.5cm} Radio Astronomy, Stanford Univ. Press, Stanford

\noindent
Jackson J C and Dodgson M 1996 MNRAS {\bf 278} 603\\
\rule{1.2cm}{0.1mm} 1997 MNRAS {\bf 285} 806\\
Kellermann K I 1993 Nat. {\bf 361} 134\\
Ozer M and Taha M O 1987 Nucl. Phys. B {\bf 287} 776\\
Pascual-Sanchez J F 1999 Preprint gr-qc/9905063\\
Peebles P J E and Ratra B 1988 Astrophys. J. {\bf 325} L17\\
Perlmutter S et al 1999 Astrophys. J. {\bf 517} 565\\
Ratra B and Peebles P J E 1988 Phys. Rev. D {\bf 37} 3406\\
Riess A G et al 1998 Astron. J. {\bf 116} 1009\\
Silveira V and Waga I 1994 Phys. Rev. D {\bf 50} 4890\\
Vilenkin A 1985 Phys. Rep. {\bf 121} 265\\
Waga I 1993 Astrophys. J. {\bf 414} 436\\
Weinberg S 1989 Rev. Mod. Phys. {\bf 61} 1\\
Zee A 1985 in High Energy Physics, Proceedings of 20th 
Annual Orbis

\hspace{.5cm} Scientiae

\end{document}